# The characterization of Co-nanoparticles supported on graphene

P. Bazylewski[1], D. W. Boukhvalov[2,6], A. I. Kukharenko[3,4], E. Z. Kurmaev[3], N. A. Skorikov[3], A. Hunt[1], A. Moewes[1], Y. H. Lee[5], S. O. Cholakh[4] and G. S. Chang[1]

[1]*Department of Physics and Engineering Physics, University of Saskatchewan, 116 Science Place,Saskatoon, SK, S7N 5E2, Canada*

[2]*Department of Chemistry, Hanyang University, 17 Haengdang-dong, Seongdong-gu, Seoul 133-791, Korea*

[3]*Institute of Metal Physics, Russian Academy of Sciences-Ural Division, 620990 Yekaterinburg, Russia*

[4]*Ural Federal University, 19 Mira Str., 620002 Yekaterinburg, Russia*

[5] *Graphene Center, Samsung Advanced Institute of Technology, Sungkyunkwan University, Suwon 440-746, Republic of Korea*

[6]*School of Computational Sciences, Korea Institute for Advanced Study (KIAS) Hoegiro 87, Dongdaemun-Gu, Seoul, 130-722, Republic of Korea*

ABSTRACT

The results of density functional theory (DFT) calculations and measurements of X-ray photoelectron (XPS) and X-ray emission (XES) spectra of Co-nanoparticles dispersed on graphene/Cu composites are presented. It is found that for 0.02nm and 0.06nm Co coverage the Co atoms form islands which are strongly oxidized under exposure at the air. For Co (2nm) coverage the upper Co-layers is oxidized whereas the lower layers contacting with graphene is in metallic state. Therefore Co (2 nm) coverage induces the formation of protective oxide layer providing the ferromagnetic properties of Co nanoparticles which can be used as spin filters in spintronics devices.

## 1. INTRODUCTION

Transition-metal nanoparticles assembled on graphene have garnered great interest in past years due to their exotic physical properties. Transition and noble-metal nanostructures on graphene have been shown to be efficient as supported catalysts. Especially, transition-metal nanoparticles (such as cobalt) on graphene has been investigated for electrochemical reduction of oxygen replacing more expensive platinum catalyst [1]. The oxygen reduction reaction takes place on the cathode side of a hydrogen fuel cell stripping electrons from hydrogen fuel at the anode and creating the electrical pull that keeps the electric current running through electrical device powered by the cell. Other technological interests are to grow uniform magnetic metal films for use as spin filters in spintronic devices and to grow high density of thermally stable magnetic islands for data storage applications. The study of metals on graphene is also important for a better understanding of the quality of metal contacts which is very critical to the performance of graphene-based electronic devices (see Ref. [2] and references therein). Magnetoelectric effects in graphene/Co systems were also predicted due to exchange coupling between cobalt clusters placed on a graphene sheet [3]. It was recently found that graphene/Cu composites exhibit a thermal conductivity higher than that of electrolytic copper which allows to use them for rapid computer cooling [4]. To understand these remarkable properties of metal/graphene composites, the full characterization of these materials on atomic and electronic level is highly desirable.

In the present paper we have used X-ray spectroscopic techniques and density functional theory (DFT) calculations to investigate the local atomic and electronic structure of Co nanoparticles assembled on graphene/Cu substrate. We have conducted X-ray photoelectron spectroscopy (XPS) measurements (Co and Cu 2*p* core levels and valence bands) and X-ray emission spectroscopy (XES) measurements (Co $L_{2,3}$ edge) of Co/graphene/Cu systems with full and partial Co coverage on the graphene surface. The obtained results are compared with DFT calculations of the formation energies for oxygen adsorption depending on the number of upper cobalt layers and the electronic structure of CoO/Co/graphene/Cu composite.

## 2. EXPERIMENTAL AND CALCULATION DETAILS

Graphene (Gr) was grown on mechanically and chemically polished Cu foil (99.96% and 100 μm-thick foil purchased from Nilaco) using atmospheric pressure chemical vapor deposition (APCVD) system (see Ref. [5] for details). The composite samples fabricated on graphene consist of a material stack of the form, Co/Gr/Cu foil with the Co layers deposited by physical vapour deposition (PVD). The thicknesses reported are as measured *in situ* by quartz crystal monitor. The prepared samples were stored and transported under light vacuum ($10^{-3}$ torr), and cleaved into smaller pieces prior to XPS measurements.

The XPS measurements at Co and Cu 2*p* core levels and valence band were made using PHI XPS Versaprobe 5000 spectrometer (ULVAC-Physical Electronics, USA). The energy resolution was $\Delta E \leq 0.5$ eV for Al *K*α excitation (1486.6 eV) and the pressure in the analysis chamber was maintained below $10^{-7}$ Pa. The dual-channel neutralizer (ULVAC-PHI patent) was applied in order to compensate the local charging of the sample due to the loss of photoelectrons during XPS measurements. All Co/Graphene/Cu samples were kept in the preparation chamber within 24 hours. After that the sample was introduced into the Analytical Chamber and controlled with the help of “Chemical State Mapping” mode in order to detect micro impurities. If the micro impurities were detected then the sample was replaced from the reserved batch. The core-level and valence-band XPS spectra were recorded in Al *K*α 100 μm spot mode with X-ray power load of the sample less than 25 W. Typical signal to noise ratio values in this case were more than 10000/3.

The XES measurements at the Co *L*-edge ($3d4s \rightarrow 2p$ transition) were conducted at the Beamline 8.0.1 of the Advanced Light Source (ALS), Berkeley, CA [6]. The incident angle of incoming photons was set at 30° to the sample surface normal with linear polarization in the horizontal scattering plane and the emitted photons from the sample were detected at an angle of 90° to the incident photons.

The pseudo-potential code SIESTA [7] was used for the DFT calculations, as done in our previous papers [8-9]. All calculations were based on the local density approximation (LDA) [10], which is feasible for the modelling of Co on graphene over Cu substrate. The full optimization of the atomic positions was performed. The wave functions were expanded with a double-ζ plus polarization basis of localized orbitals for all species. The force and total energy of system were optimized with an accuracy of 0.04 eV/Å and 1 meV, respectively. All calculations were carried out with an energy mesh cut-off of 360 Ry and a **k**-point mesh of 4×4×2 in the Monkhorst-Pack scheme [11]. The chemisorption energies for oxidation was calculated using a standard equation: $E_{chem} = E_{host+O} - (E_{host} + E_{O2})$ where $E_{host}$ is the total energy of system before adsorption of oxygen atom and $E_{O2}$ is the total energy of molecular oxygen in triplet state.

The oxidation of Co/Gr/Cu system is simulated using the following model structures: (*i*) Co clusters of 1, 2 or 7 atoms on the graphene/Cu substrate consisting of 6 *fcc* Cu layers (16 Cu atoms in each layer) covered with a layer of graphene containing 32 C atoms (6, 11 or 43% of the Co coverage) and (*ii*) 1, 2, 4, or 6 layers of Co on the same graphene/Cu structure. A simulation of the oxidation is produced by placing the oxygen molecule close to the Co atoms, which is then decomposed at the surface by forming bonds with Co atoms. Figures 1 (a) and (b) represent the model graphene/Cu structures with 6-atom Co cluster and 6 Co layers, respectively.

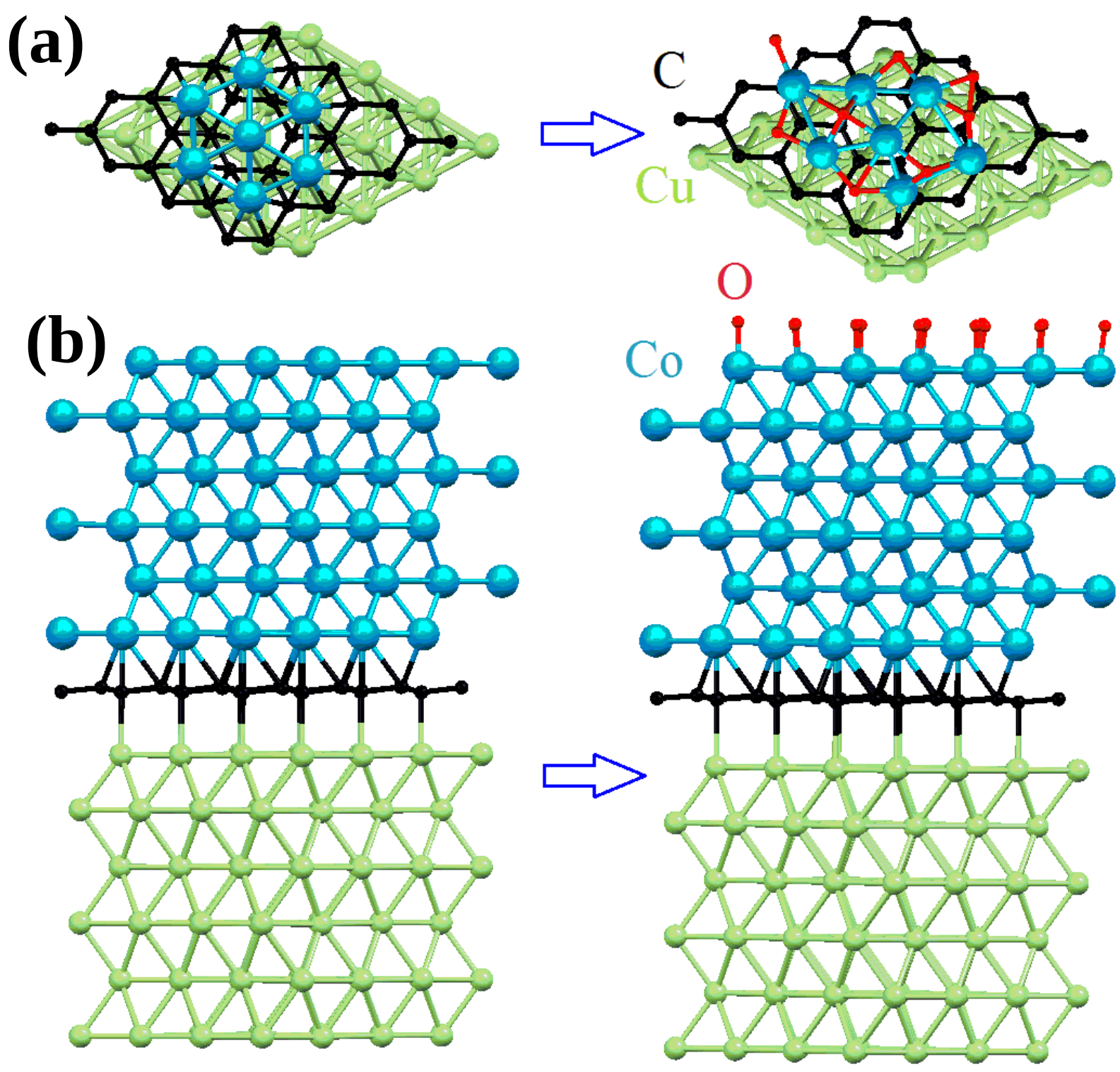

FIG. 1. Optimized atomic structure for Co cluster covering half of graphene surface on Cu-substrate with 6 upper Co-layers.

Magnetic and spectral behaviors of CoO/Co/Gr/Cu composites were investigated using the pseudo-potential Quantum-Espresso code with Perdew-Wang (PW) version of exchange potential [10,16]. For the simulation of the CoO/Co/Gr/Cu composites, the following model structures were considered: 1 or 2 layers of *fcc* Co and 1 layer of O atoms were placed on the graphene/Cu substrate as above described (i). The 10 Å of empty space was added on top of oxygen layer. The atomic positions in the constructed cells were relaxed while each component of force was more than 2 mRy/a.u. The relaxation and final calculations were performed with a **k**-point mesh of 6x6x1 in the Monkhorst-Pack scheme [11]. The plane-wave and kinetic energy cut-offs were chosen to be 40 Ry and 200 Ry, respectively.

## 3. RESULTS AND DISCUSSION

The XPS survey spectra of Co-coated graphene/Cu composites normalized to the Cu 2$p_{3/2}$ XPS line of Cu foil are presented in Fig. 2. The XPS spectra show no additional impurities presented in all Co/Gr/Cu samples and the intensity of Co 2*p* line increases with increasing the Co content. The chemical state of Co supported atoms can be determined with help of measurements of Co 2$p_{3/2,1/2}$ core level XPS lines.

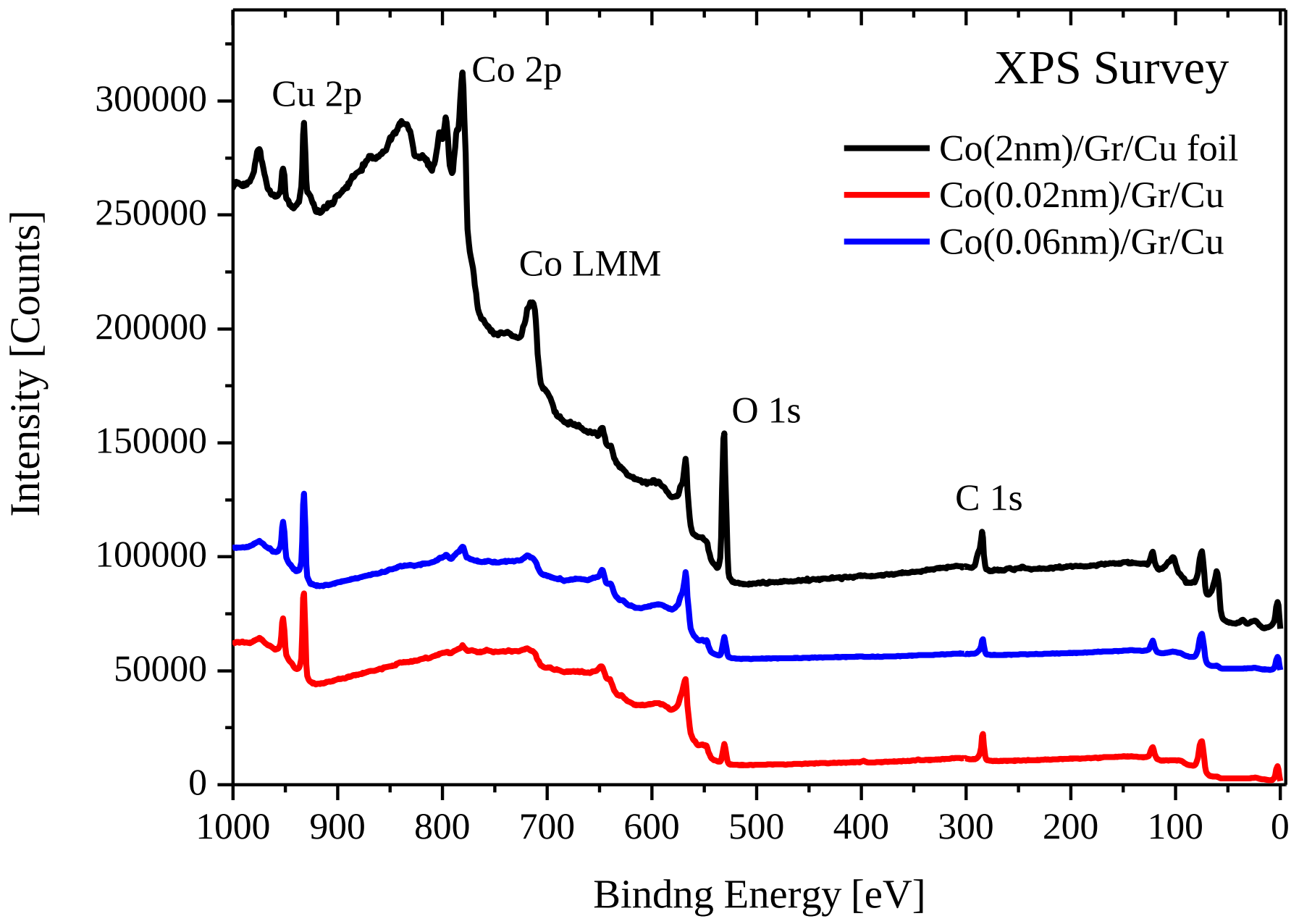

FIG. 2. XPS survey spectra of Co(0.02nm)/Gr/Cu, Co(0.06nm)/Gr/Cu and Co(2nm)/Gr/Cu (spectra are normalized for Cu 2*p*).

As seen in Fig. 3, Co 2*p* XPS spectrum of Co(2nm)/Gr/Cu has a rather complicated structure which is attributed to superposition of $Co^0$ and $Co^{2+}$ signals in ratio 30:70 (see Fig. 4). On the other hand, the spectra of Co islands (0.02nm Co/Gr/Cu and 0.06nm Co/Gr/Cu) are relatively simple and similar to that of $Co(OH)_2$ with charge transfer satellites which are typical for spectra of divalent Co ($Co^{2+}$) [12].

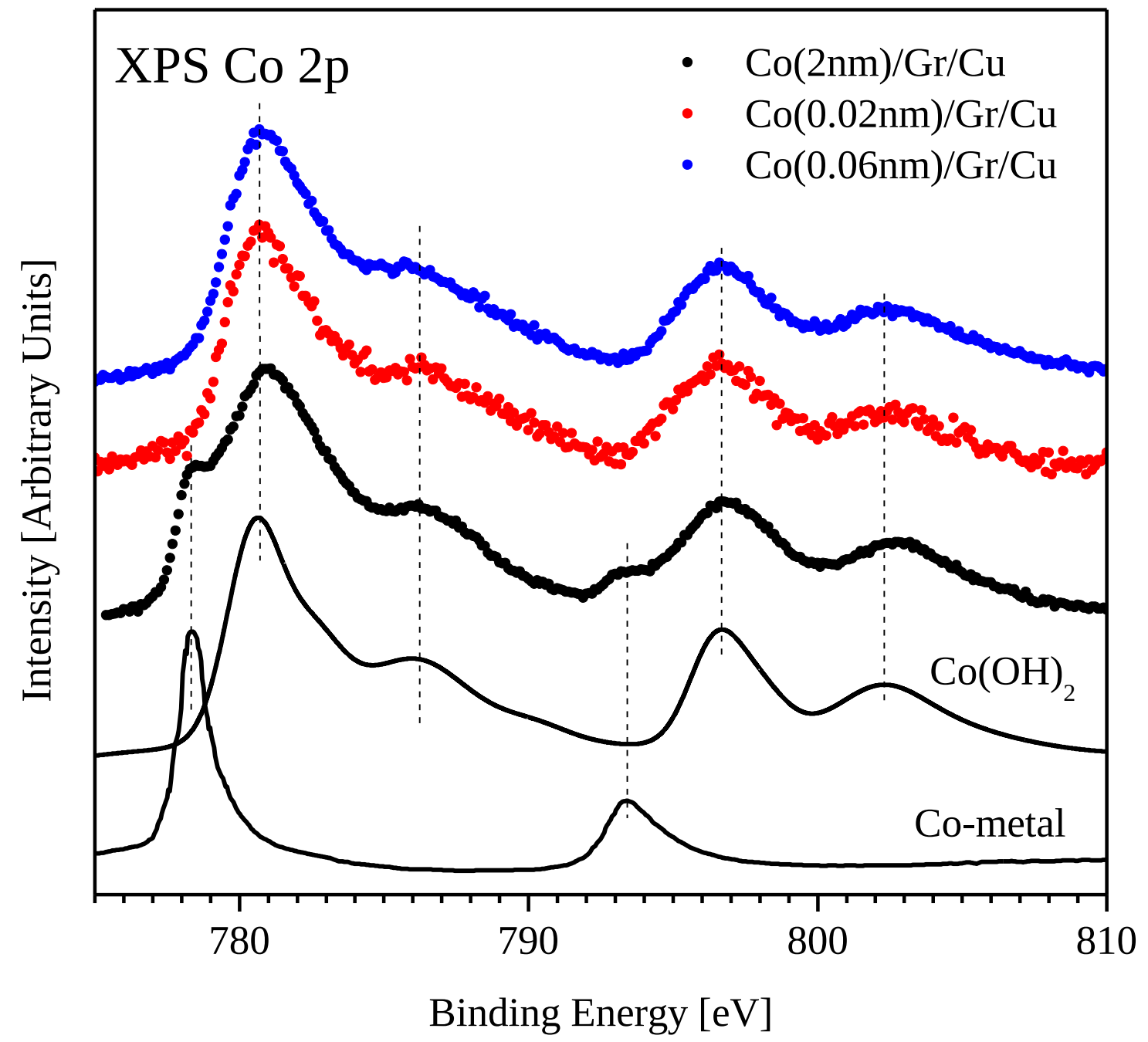

FIG. 3. The comparison of Co 2*p* XPS spectra of Co-coated graphene/Cu systems with spectra of reference samples (Co metal and $Co(OH)_2$ [6]).

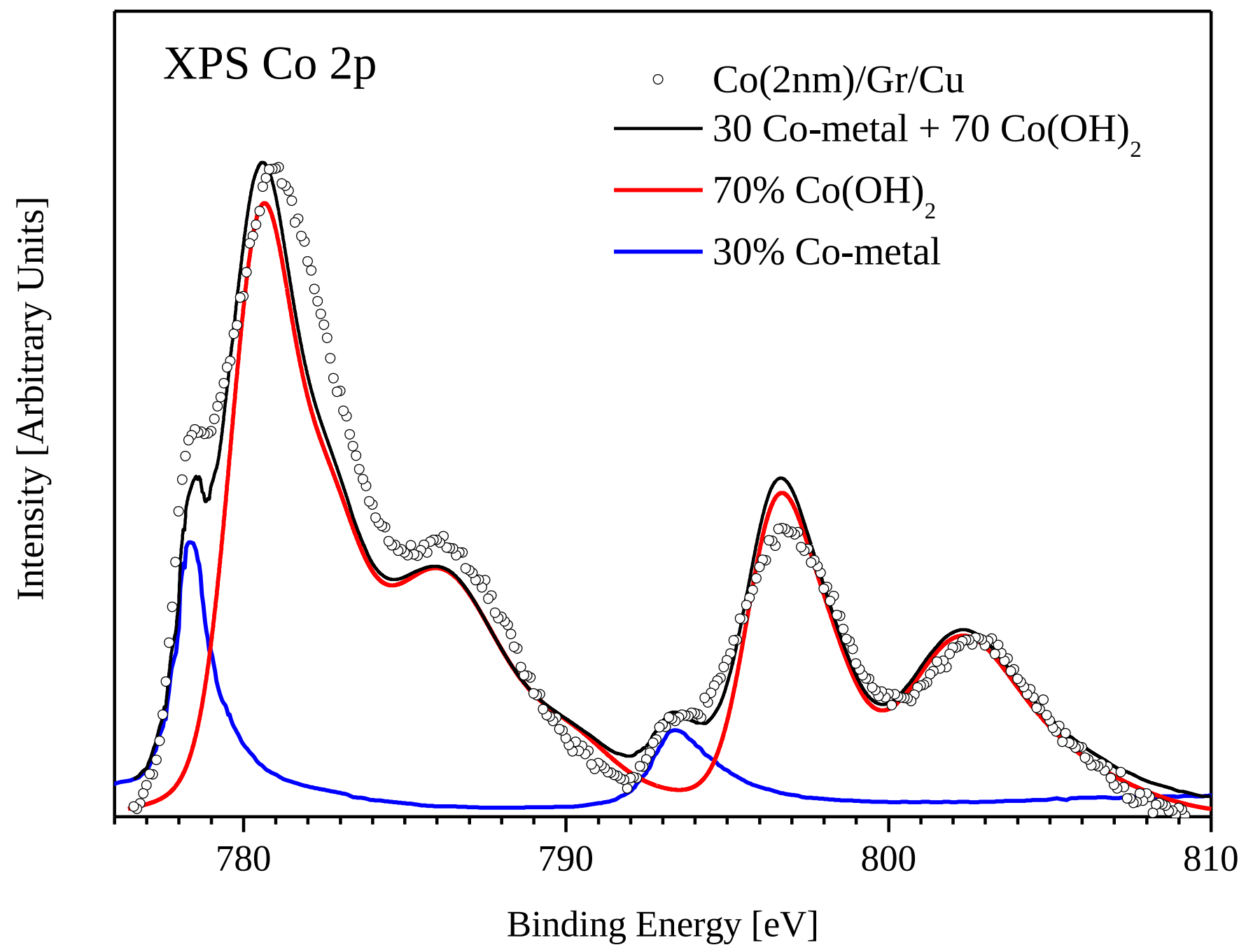

FIG. 4. The comparison of Co 2*p* XPS spectrum of Co(2nm)/Gr/Cu with a superposed spectrum with $Co(OH)_2$ and Co metal reference spectra.

This divalency of Co in the Co-dispersed graphene/Cu samples is consistent with results of non-resonant Co $L_{2,3}$ XES for (0.02 nm)Co/Gr/Cu composite as presented in Fig. 5. One can see that the relative intensity ratio of $L_2$ emission line to $L_3$ line, I($L_2$)/I($L_3$), for (0.02 nm)Co/Gr/Cu sample is almost the same as that of CoO and much higher than that of Co metal. The I($L_2$)/I($L_3$) intensity ratio is related with the probability of radiationless $L_2L_3M_{4,5}$ Coster–Kronig (*C*–*K*) transitions and the ratio of total photo-absorption coefficients ($\mu_3/\mu_2$) for excitation energies at the $L_2$ and $L_3$ absorption thresholds [13]. Since the $\mu_3/\mu_2$ ratio depends only on the excitation energy, the I($L_2$)/I($L_3$) intensity ratio of Co *L*-edge XES spectra taken at the same excitation energy is determined by the *C*–*K* transitions, which are governed by the number of free *d*-electrons around a target atom. Therefore, the results of non-resonant Co $L_{2,3}$ XES measurements confirm the presence of $Co^{2+}$ species in 10 Co/Gr/Cu sample.

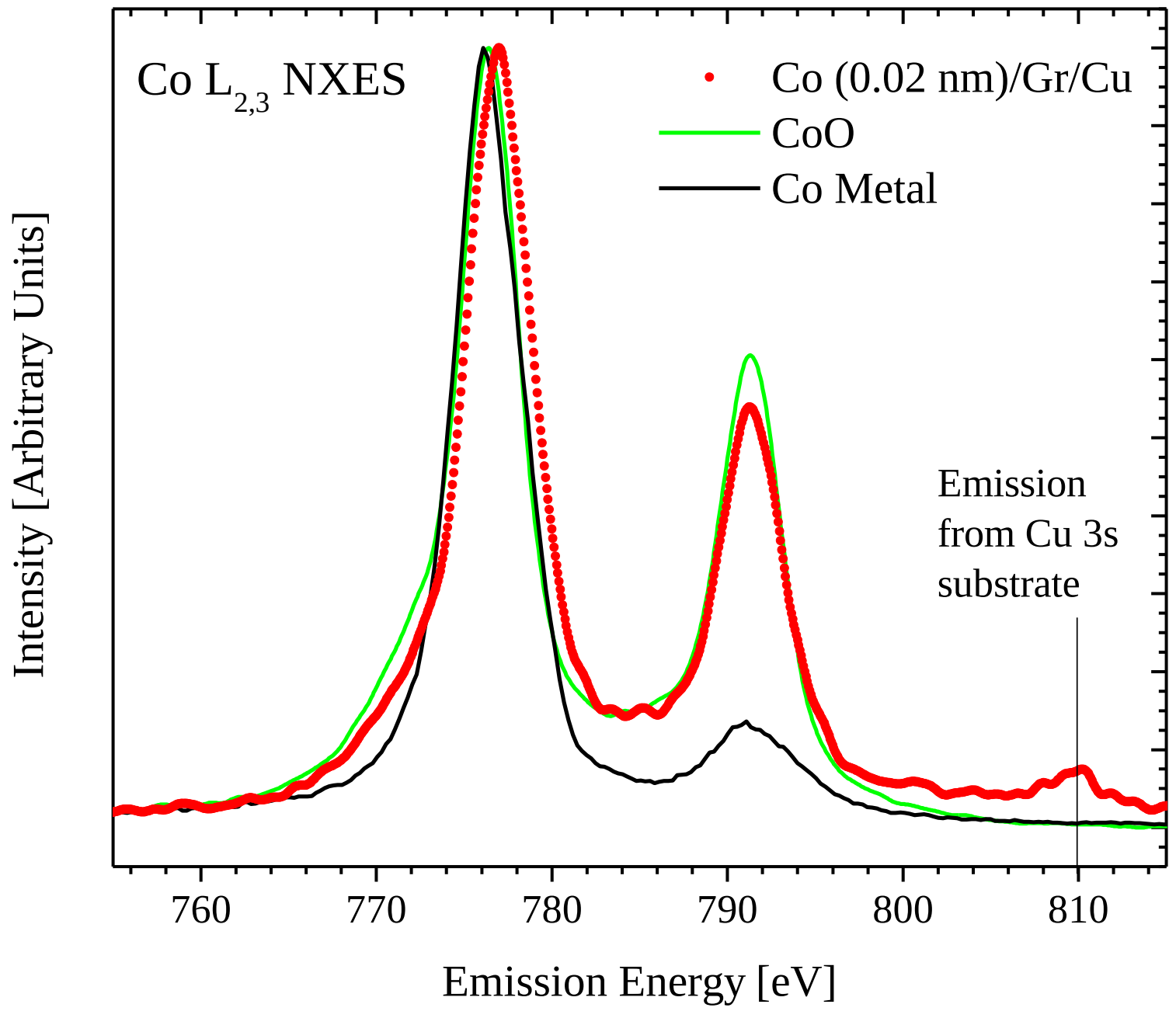

FIG. 5. Non-resonant Co $L_{2,3}$ XES of (0.02 nm) Co/Gr/Cu composite and reference samples (CoO and Co metal)

On the other hand, Cu 2*p* XPS spectra of Co/Gr/Cu composites in Fig. 6 show similar spectral behavior to that of Cu-metal, suggesting that polished Cu-foil is not oxidized after graphene synthesis and Co deposition. Figure 7 presents the valence band XPS spectra of 0.02nm Co/Gr/Cu, 0.06nm Co/Gr/Cu and Co(2nm)/Gr/Cu. The spectral feature at about 2.6 eV is mainly contributed from Cu *3d* states as typical for Cu metal [14] because of high photoionization cross-section of Cu 3*d* states for Al *K*α radiation compared to those of C 2*p* and O 2*p* states [8]. For Co(2nm)/Gr/Cu sample, the Co 3*d* contribution from Co metal [14] and CoO [9] at about 0.7 eV becomes more pronounced than those of (0.02nm) Co/Gr/Cu and (0.06nm) Co/Gr/Cu due to higher Co content.

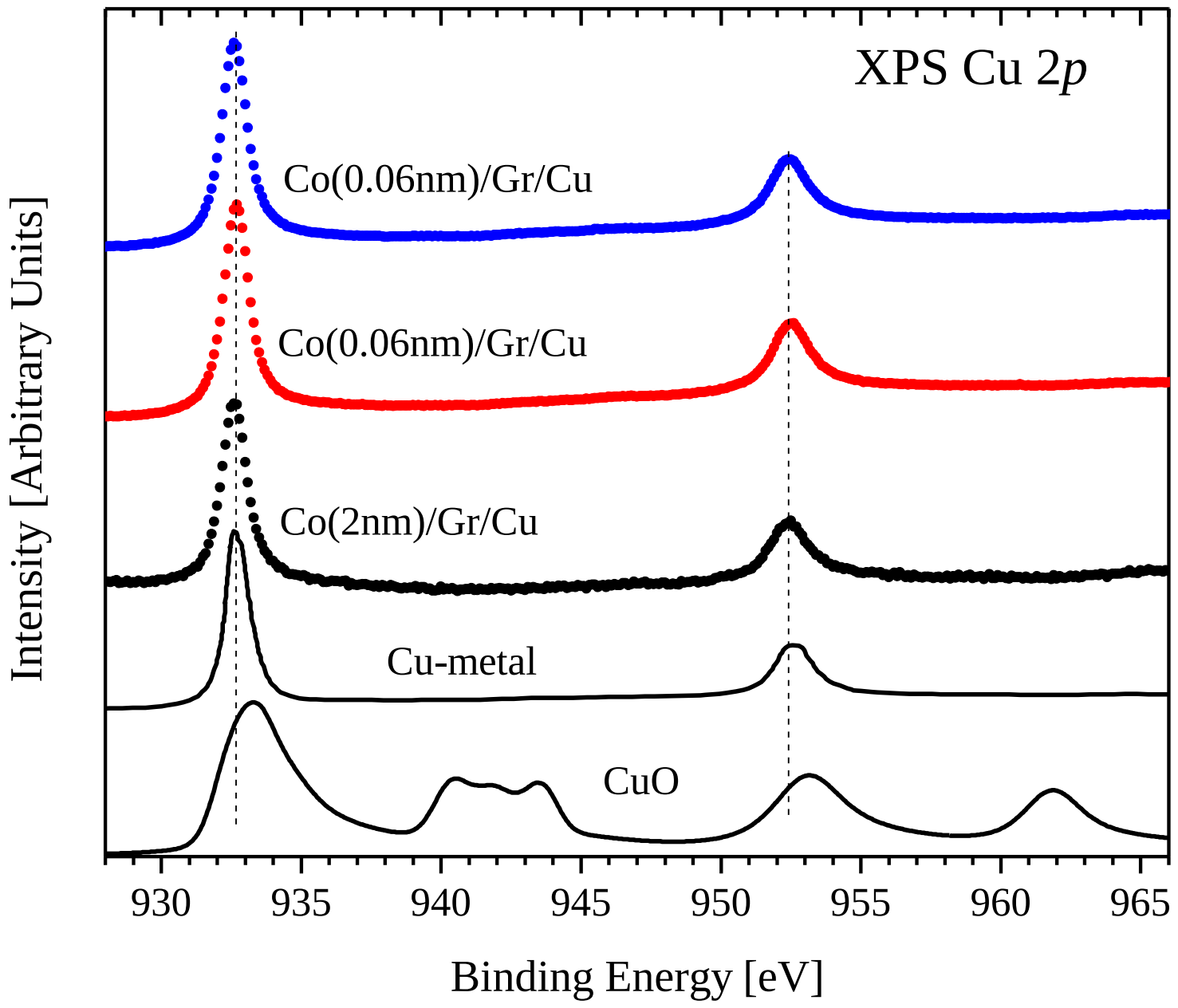

FIG. 6. Cu 2*p* core level XPS spectra of graphene/Cu substrates with different Co coverages.

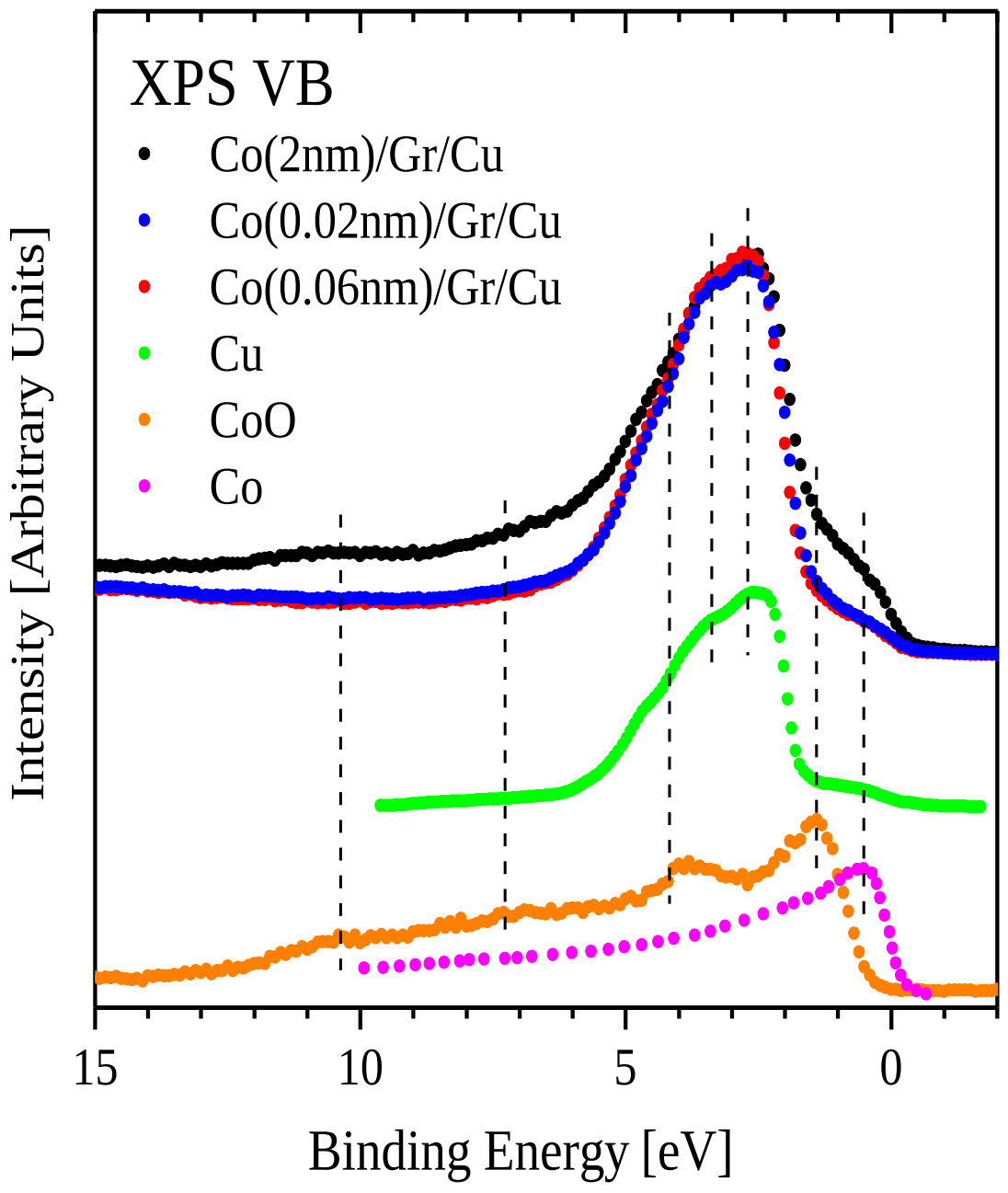

FIG. 7. The comparison of valence band (VB) XPS spectra of (0.02nm) Co/Gr/Cu, (0.06nm) Co/Gr/Cu and Co(2nm)/Gr/Cu with reference spectra of Cu metal [14], Co metal [14] and CoO [15].

The valence band XPS spectra are in a good agreement with DFT calculations of the electronic structure of CoO/Co/Gr/Cu composite. The calculation results are presented in Fig. 8. The contribution of partial density of states (PDOS) of Co metal atoms (Co1) dominates near the Fermi level whereas contributions of PDOS of Co(2) and O atoms in the upper CoO layer are located at lower part of the valence band. The spin magnetic moments of Co ions are found to be ferromagnetically arranged in the case of two layers of Co. The magnetic moments values are found to be different between CoO and Co layers: 0.3 $\mu_B$/atom for CoO layer and 1.4 $\mu_B$/atom for Co layer closest to the graphene. For a model cell containing one layer of Co, we get nonmagnetic solution.

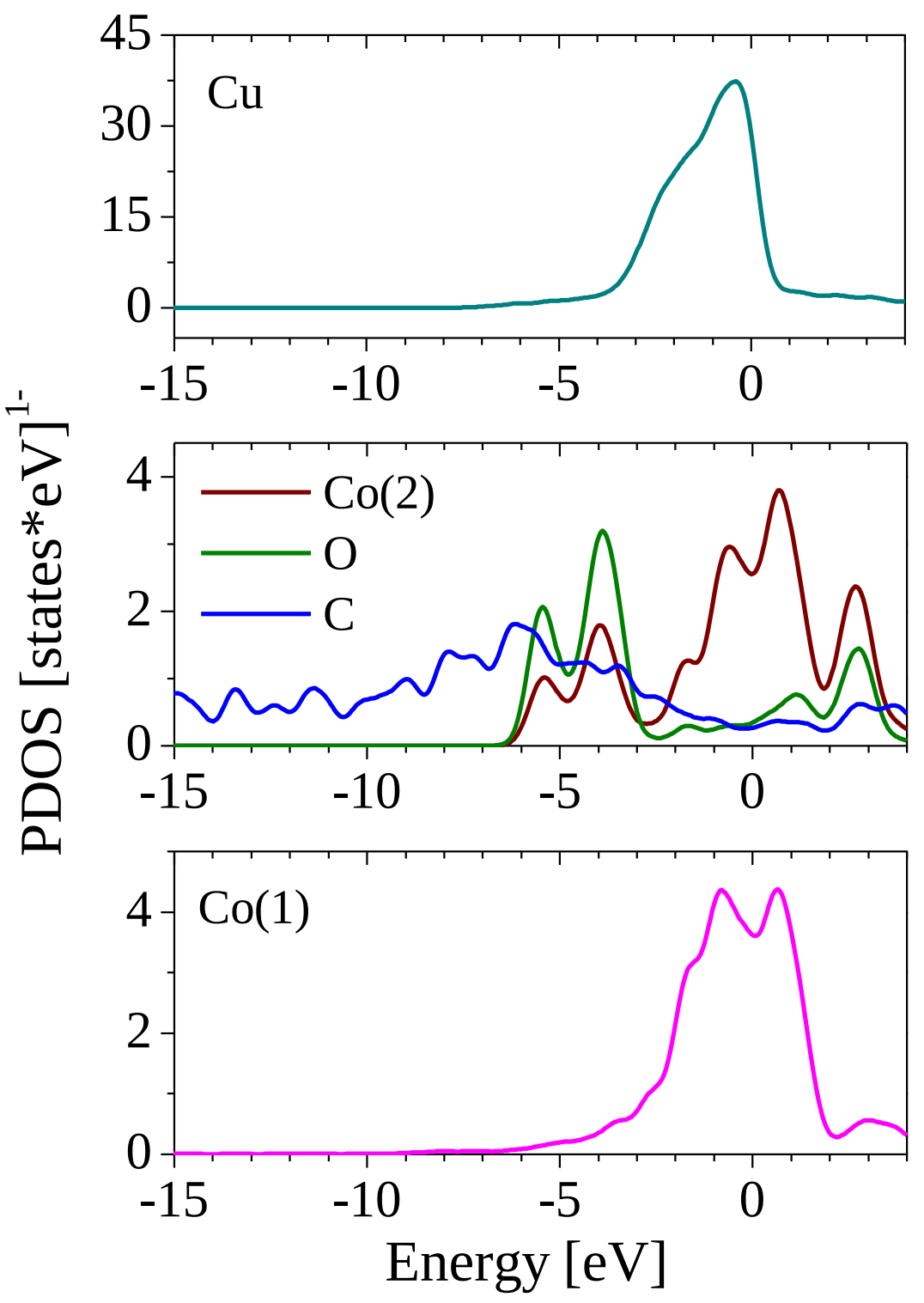

FIG. 8. Calculated partial densities of states for CoO/Co/graphene/Cu composite.

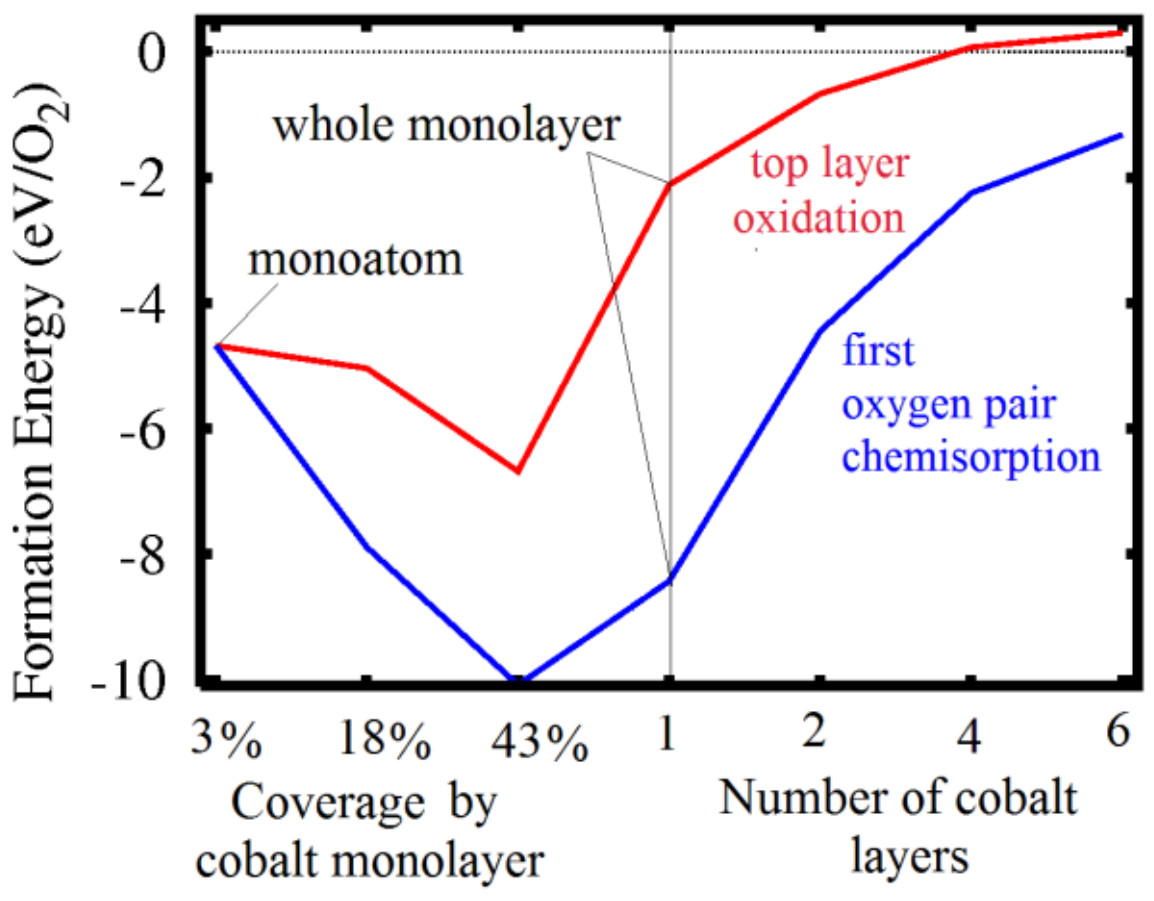

FIG. 9. Calculated formation energies of adsorbed oxygen atoms with respect to the degree of Co coverage and the number of Co layers.

Finally, the formation energies of adsorbed oxygen atoms in dependence of the degree of Co coverage and the number of Co layers are presented in Fig. 9. It is clearly seen that with the growth of the Co clusters, the oxidation process becomes energetically favorable as well as full oxidation of the cluster. In contrast to the Co monolayer, the Co clusters can easily change their morphology under oxidation and move to 3-dimensional systems such as Co oxides (see Fig. 1a). The oxidation of Co mono-atom is less energetically favorable than Co cluster because after addition of oxygen atoms it turns into $CoO_2$ molecule on graphene. With increasing number of Co layers the joining of the first oxygen molecule becomes more and more energetically difficult, but still remains an exothermic process. Further full oxidation of the top Co layer with increasing Co thickness up to 1 nm, the process becomes endothermic, but with small formation energy, the process can be realized at room temperature. This suggests that the process of oxidation of Co thin films on graphene/Cu substrate will be rather slow and eventually stop. As shown in Fig. 4, the ratio of oxidized Co to metallic Co in Co(2nm)/Gr/Cu sample is close to 7:3. However, it should be noted that Co oxidation takes place not only on the surface but also at the grain boundaries of Co while the DFT calculations do not take into account. In this case, the volume of Co remains unoxidized which may be explained by a combination of the following effects:

1) The presence of graphene substrate reduces the gaps between the Co grains because the adsorption of Co on graphene is determined by morphology of substrate.
2) Additional calculations for the oxidation of 1, 2, 4 and 6 layers of Co without Gr/Cu substrate demonstrate that oxidation of Co layers without Gr/Cu is energetically more favorable (0.2~0.4 eV/$O_2$). Therefore, the presence of Gr/Cu substrate decreases chemical activity of all sites of Co.

## 4. **CONCLUSION**

To conclude, we have investigated a series of Co/graphene/Cu composites with tunable Co nanoparticle coverage. Based on results of DFT calculations and spectroscopic measurements using XPS and XES, we conclude that at the 0.02nm and 0.06nm of Co coverage the Co atoms are located on the surface of graphene/Cu substrate as islands. These Co islands are immediately oxidized in ambient air. For the Co thickness of 2 nm, the upper Co layers (70% of full thickness) are oxidized whereas the lower layers (30%) are still in metallic state. For two or more layers of Co atoms beneath an oxidized CoO layer the ferromagnetic behavior is theoretically predicted. Therefore the formation of protective oxide layer prevents Co from deep oxidation and allows to use its ferromagnetic properties for spintronics applications.

**Acnowledgements**

The results of experimental measurements of XPS spectra were obtained by A.I. Kukharenko, E.Z. Kurmaev and S.O. Cholakh and supported by the grant of the Russian Scientific Foundation (Project No. 14-22-00004). Theoretical calculations of electronic structure of Cu/Gr/Co/CoO composite were performed by N.A. Skorikov and supported by the grant of the Russian Scientific Foundation (Project No. 14-22-00004). G.S. Chang gratefully acknowledges support from the Natural Sciences and Engineering Research Council of Canada (NSERC) and Canada Foundation for Innovation (CFI).